\documentclass[twocolumn, showkeys]{IEEEtran}
\usepackage{epsfig}
\usepackage{graphicx}% Include figure files
\usepackage{amsmath}% Include amsmath package
\usepackage{dcolumn}% Align table columns on decimal point
\usepackage{bm}% bold math
\usepackage{algpseudocode, algorithm}
\usepackage{tikz}
\usepackage[all]{xy}
\usetikzlibrary{arrows,shapes}
\usepackage{pgf-umlsd}
\usepackage{url}

\usepackage{centernot}
\usepackage{stmaryrd}

\usepackage{color} % REMOVE BEFORE SUBMITTING

\usepackage{authblk}
\usepackage{amsmath}
\usepackage{amssymb}
\usepackage{float}

\def\BibTeX{{\rm B\kern-.05em{\sc i\kern-.025em b}\kern-.08em T\kern-.1667em\lower.7ex\hbox{E}\kern-.125emX}}

\begin{document}

\title{Wireless Resilient Routing Reconfiguration}

\author[1]{Brian DeCleene}
\author[2]{Steve Huntsman}
\affil[1]{BAE Systems FAST Labs\\
600 District Avenue, Burlington, MA 01803\\
brian.decleene@baesystems.com}
\affil[2]{BAE Systems FAST Labs\\
4301 Fairfax Drive, Arlington, VA 22201\\
steve.huntsman@baesystems.com
\vspace*{-1cm}}

\date{\today}

\IEEEoverridecommandlockouts
\IEEEpubid{\makebox[\columnwidth]{\copyright2019 BAE Systems, Inc. \hfill} \hspace{\columnsep}\makebox[\columnwidth]{ }}

\maketitle

\thispagestyle{empty}

\begin{abstract}
Mobile wireless networks are intrinsically more prone to link congestion and outright failures than wired networks. In this paper, we elaborate the resilient routing reconfiguration method of \cite{WangEtAl} and generalize it to accomodate point-to-multipoint links and wireless networks. By reframing link failures as traffic uncertainties, this technique allows essentially instantaneous rerouting around arbitrary link failures while preventing congestion. We illustrate the technique by identifying a critical bottleneck in a realistic model wireless network.
\end{abstract}


\section{\label{sec:Introduction}Introduction}

It is desirable for networks to be resilient in the face of link failures. However, naive methods for generating routing protection schemes that account for congestion have complexity that grows combinatorially with the number of failures to protect against. That is, if a network has $N$ links and protection is required for up to $F$ link failures, then there are $\sum_{k=0}^F \binom{N}{k}$ possible failure scenarios to plan for. For $N = 100$ and $F = 2$, this number is $5051$; for $F = 4$, it is $4087976$. This scaling behavior precludes brute-force approaches to resilient traffic engineering. Furthermore, planning for the combinatorally large number of scenarios should be coordinated in such a way to minimize disruptions to the traffic pattern when new failures occur. In short, optimizing traffic routing with respect to individual failure scenarios is an inadequate traffic engineering strategy. 

An approach that overcomes the problems above was presented in \cite{WangEtAl} using \emph{R3}, a congestion-avoiding routing reconfiguration framework that is resilient under multiple failures. The basic idea behind R3 is to account for all possible failure scenarios within a single optimization problem by adding ``virtual'' traffic corresponding to the capacity of links that might fail. This converts uncertainty in network topology into uncertainty in traffic. A \emph{base routing} $r$ that optimizes maximum link utilization is solved for along with a \emph{protection routing} $p$ that encodes detours in the presence of link failures. As links fail, $r$ and $p$ are updated using a handful of simple arithmetic operations, and traffic is rerouted accordingly. The simplicity of the updates minimizes network losses and latency. Meanwhile, in the background a solver can continuously monitor the current network connectivity and solve for optimal base and protection routings to replace the near-optimal updates as network stability permits.

R3 enjoys theoretical guarantees regarding congestion avoidance, optimality, and the order of link failures. It is also efficient in practice, where the theoretical requirements for these guarantees do not typically hold. For example, a single node on the network periphery may be isolated with fewer than $F$ failures, but the traffic pattern that R3 generates will not be adversely affected by this degeneracy. In this work, we detail R3 \emph{en route} to adapting it to (mobile) wireless networks. 

Proactive alternatives to R3 were proposed in \cite{SucharaEtAl} and \cite{LiuEtAl}: however, their reliance on predicted traffic demand adds an element of uncertainty that R3 avoids, while \cite{LiuEtAl} also focuses on node failures versus link failures, making it less relevant for mobile networks. Furthermore, these alternatives do not offer the theoretical guarantees of R3 (cf. \cite{Chiesa1,Chiesa2,Chiesa3} for general theoretical considerations of routing resiliency). The approaches of \cite{Foerster,Foerster2} incorporate route quality (e.g., length and congestion) as well as connectivity into fast failover routing, but the underlying mathematical problems are $\mathbf{NP}$-hard, hindering general adaptation to mobile wireless networks. Other approaches to fast rerouting/failover such as \cite{Pignolet,Stephens} have also not yet been extended to the wireless domain: to our knowledge, this paper details the first resilient traffic engineering technique suitable for wireless networks.

We proceed in \S \ref{sec:Informal} with a brief informal discussion to make this paper relatively self-contained. In \S \ref{sec:BasicRoutingConstraints} we establish more formal notations and definitions for the basic quantities of interest to R3, and we demonstrate how basic routing constraints can be effectively formulated using tensor product structure. We continue this approach in \S \ref{sec:R3LP}, conveniently giving the linear program embodying the offline precomputation for R3 in explicit matrix form. We address technicalities arising in the adaptation of R3 to wireless networks in \S \ref{sec:Technicalities}. In \S \ref{sec:WirelessR3} we move on to understand point-to-multipoint communications of the sort prevalent in wireless networks before introducing the corresponding generalization of R3 in \S \ref{sec:WR3LP}.

\section{\label{sec:Informal}Informal overview of R3}

Let $G = (V,L)$ be a directed multigraph modeling the network topology: network routers correspond to vertices in $V$, and network links are represented by directed edges in $L$. 
It will be convenient to write $s(\ell)$ and $t(\ell)$ respectively for the source and target (sink) of a link $\ell$. 
Let $d : V^2 \rightarrow \mathbb{R}_{\ge 0}$ with zero diagonal (i.e., $d(a,a) \equiv 0$) be the \emph{traffic demand} and write $d_{ab} := d(a,b)$. Let $c : L \rightarrow \mathbb{R}_{\ge 0}$ be the \emph{link capacity}. 
If $a,b \in V$ and $\ell \in L$, then the value $r_{ab}(\ell)$ of a base routing $r$ specifies the fraction of traffic with origin $a$ and destination $b$ that traverses the link $\ell$. Thus the total amount of traffic on link $\ell$ is $\sum_{a,b \in V} d_{ab} r_{ab}(\ell)$. More generally, a routing (defined formally in \eqref{eq:routing} below) is any function from $V^2 \times L$ to $[0,1]$ that satisfies natural constraints corresponding to conservation, totality, and global acyclicity of flow. 

In the R3 framework, the capacitated topology $(V,L,c)$ and demand $d$ are given along with a number $F$ of allowed link failures. A base routing $r$ and protection routing $p$ are derived to ensure congestion-free traffic flow under $\le F$ link failures if sufficient connectivity exists. 

The protection routing $p$ has the particular requirement that its nontrivial origin/destination pairs are of the form $(s(\ell),t(\ell))$, and it encodes weighted alternative paths from $s(\ell)$ to $t(\ell)$. Thus when link $\ell$ fails, the remaining paths from $s(\ell)$ to $t(\ell)$ can be reweighted and used in place of $\ell$. This reconfiguration (which applies to both $r$ and $p$) only requires simple arithmetic operations and can be applied essentially instantaneously once a link failure is detected. Meanwhile, a background process can continuously solve for base and protection routings for the current topology and number of remaining allowed link failures to smoothly transition from optimal pre-planned routes to routes that are optimal for the actual current failures and residual possible failures.

To plan for arbitrary link failures, we use the \emph{rerouting virtual demand set} $Z_F := \{z : (0 \le z \le c) \land (\sum_{\ell'} z_{\ell'}/c_{\ell'} \le F) \}$. Each point $z \in Z_F$ corresponds to a potential load on the network that saturates no more than $F$ links on its own. In principle $r$ and $p$ could be obtained by solving the constrained optimization problem
\begin{align}
\label{eq:R3LPprecursor}
\min_{r,p} \mu \quad \text{s.t. $r$ and $p$ are routings};
	& \\
\underbrace{\sum_{a,b \in V} d_{ab} r_{ab}(\ell)}_{\text{actual traffic on link } \ell} + \underbrace{\max_{z \in Z_F} \sum_{\ell' \in L} z_{\ell'} p_{s(\ell'),t(\ell')}(\ell)}_{\text{maximum virtual traffic on link } \ell} \quad \overset{\forall \ell \in E}{\le}
	& \quad c_\ell \mu. \nonumber 
\end{align}
This optimization requires that the sum of actual and maximum virtual traffic not exceed the link capacity times the maximum link utilization $\mu$. So long as the objective $\mu \le 1$, congestion-free routing is possible under $\le F$ link failures (and frequently in practice this works nicely even if $F$ failures can partition the network, since the online reconfiguration can remove unreachable demands).

In practice, the form of the optimization problem above is not immediately useful. However, it can be transformed into an equivalent linear program using the duality theorem. We elaborate on this transformation and the actual linear program we work with in \S \ref{sec:R3LP}. The solution time varies only indirectly with $F$, though for larger values more redundancy is demanded of a solution and routing performance will necessarily be affected. Thus the value of $F$ chosen should reflect some specific planning consideration.

With $r$ and $p$ in hand, traffic can be routed using $r$ and reconfigured using both $r$ and $p$ as follows. If link $\ell$ fails, we reconfigure $(r,p) \mapsto (r',p')$ according to
\begin{equation}
\label{eq:update_r0}
r'_{ab}(\ell') := r_{ab}(\ell') + r_{ab}(\ell) \cdot \xi_\ell (\ell');
\end{equation}
\begin{equation}
\label{eq:update_p0}
p'_{s(\ell')t(\ell')}(\ell'') := p_{s(\ell')t(\ell')}(\ell'') + p_{s(\ell')t(\ell')}(\ell) \cdot \xi_\ell (\ell''),
\end{equation}
where
\begin{equation}
\label{eq:xi}
\xi_\ell (\ell') := \begin{cases}
0 & \text{ if } p_{s(\ell)t(\ell)}(\ell) = 1; \\
\frac{p_{s(\ell)t(\ell)}(\ell')}{1-p_{s(\ell)t(\ell)}(\ell)} & \text{ otherwise}.
\end{cases}
\end{equation}
This simple update rule is also applied for subsequent failures and yields essentially instantaneous rerouting. %Details are in \S \ref{sec:R3Online}.

There are three major subtleties in the offline configuration phase of R3 in which the base routing $r$ and protection routing $p$ are computed that are not addressed in \cite{WangEtAl}. The first of these subtleties is the intricate indexing required in setting up the key linear program. The second and third are related to parallel links and the preservation of routing constraints. These are respectively tackled by judicious use of tensor algebra in \S \ref{sec:BasicRoutingConstraints} and \S \ref{sec:R3LP}, a topology virtualization step that uses virtual nodes to eliminate parallel links (necessary for the self-consistency of the framework) combined with load evaluation as detailed in \S \ref{sec:Parallel}, and auxiliary techniques as mentioned in \S \ref{sec:RoutingPreservation}.

Finally, R3 was developed for wired network backbones: however, we have extended the approach in such a way that it can apply to networks with both wired and wireless connections. The key is to impose an additional constraint that ties the capacity of a wireless transmitter to a point-to-multipoint connection incorporating multiple links.

\section{\label{sec:BasicRoutingConstraints}Basic routing constraint}

A function $r : V^2 \times L \rightarrow [0,1]$, written $r((a,b),\ell) =: r_{ab}(\ell)$, is called a \emph{(flow representation of a) routing} if the following conditions are satisfied for all $(a,b,\ell) \in V^2 \times L$:
\begin{subequations}
\label{eq:routing}
\begin{align}
r_{aa}(\ell) \quad =
	& \quad 0; \label{eq:RoutingSelf} \\
\sum_{\ell: s(\ell) = j} r_{ab}(\ell) \quad =
	& \quad \sum_{\ell' : t(\ell') = j} r_{ab}(\ell'), \quad *; \label{eq:RoutingConservation} \\
	% & \quad \sum_{\ell' : t(\ell') = j} r_{ab}(\ell'), \quad a,b,j \text{ distinct}, \quad j \text{ not a source or target of } G; \label{eq:RoutingConservation} \\
\sum_{\ell : s(\ell) = a} r_{ab}(\ell) \quad = 
	& \quad 1, \quad \text{$a \ne b$, $a$ not a target of $G$}; \label{eq:RoutingOutTotality} \\
\sum_{\ell: t(\ell) = b} r_{ab}(\ell) \quad = 
	& \quad 1, \quad \text{$a \ne b$, $b$ not a target of $G$}; \label{eq:RoutingInTotality} \\
r_{ab}(\ell) \quad =
	& \quad 0, \quad \text{$a \ne b$, $t(\ell) = a$}; \label{eq:RoutingNoReturn} \\
r_{ab}(\ell) \quad =
	& \quad 0, \quad \text{$a \ne b$, $s(\ell) = b$}. \label{eq:RoutingNoExtension}
\end{align}
\end{subequations}
Here $*$ in \eqref{eq:RoutingConservation} indicates that $a,b,j$ are all distinct, and that $j$ is neither a source nor a target of $G$.

We note that \cite{WangEtAl} ignores the requirement in \eqref{eq:RoutingConservation} that $j$ should not be a source or target of $G$ [i.e., that $j$ should have positive in- and out-degrees], omits \eqref{eq:RoutingInTotality} and \eqref{eq:RoutingNoExtension}, and notationally suggests that there are no parallel links: however, all of these modifications are self-evidently desirable, not least in that they avoid degeneracies and manifestly enforce symmetry. That said, it may be desirable for the sake of computational efficiency to omit \eqref{eq:RoutingInTotality} and \eqref{eq:RoutingNoExtension}.

It turns out to be useful to deal with a weaker notion than a routing. For instance, a routing for the graph in Figure \ref{fig:ThreeNodes} must take spurious nonzero values.
\begin{figure}
	\begin{minipage}[c]{0.75\columnwidth}
		\caption{
			\label{fig:ThreeNodes}A routing $r$ on this graph behaves poorly, e.g. $r_{ab}(\ell) = 1$ for $(a,b,\ell) \in \{(2,3,2),(3,1,3)\}$, despite the fact that in both cases there is no path from $a$ to $b$, much less one traversing $\ell$ with $t(\ell) \ne a$ and $s(\ell) \ne b$. 
	    	}
  	\end{minipage}
	\begin{minipage}[c]{0.2\columnwidth}
		\begin{tikzpicture}[scale=.75,->,>=stealth',shorten >=1pt,every node/.style={transform shape}]
			\node [draw,circle] (v1) at (0,0) {1};
			\node [draw,circle] (v2) at (2,0) {2};
			\node [draw,circle] (v3) at (1,1) {3};
			\path [->] (v1) edge node [below] {$1$} (v2);
			\path [->] (v1) edge node [above] {$2$} (v3);
			\path [->] (v3) edge node [above] {$3$} (v2);
		\end{tikzpicture}
	\end{minipage}\hfill
\end{figure}
Although in most respects such spurious values are harmless, they also involve equations to pointlessly solve and they complicate our understanding. As such we mention the weaker notion of a \emph{semirouting}, in which \eqref{eq:routing} is satisfied only for $(a,b,\ell) \in V^2 \times L$ such that there are paths in $G$ from $a$ to $s(\ell)$ and from $t(\ell)$ to $b$, and such that $t(\ell) \ne a$ and $s(\ell) \ne b$. A \emph{restricted semirouting} that identically takes the value zero on triples not of this form is also useful to consider. That said, we restrict ourselves to routings in the rest of this paper.

Much of the effort in setting up a more useful equivalent of \eqref{eq:R3LPprecursor} is tied to intricate indexing that some basic tensor algebra can clarify. Without loss of generality, let $V = [n] \equiv \{1,\dots,n\}$ and $L = [N]$, so that $|V| = n$ and $|L| = N$. Let ${\bf e}^{(n)}_j$ denote the $j$th standard basis vector in $\mathbb{R}^n$: then ${\bf e}^{(n)}_j \otimes {\bf e}^{(n')}_{j'} = {\bf e}^{(nn')}_{(j-1)n'+j'}$, where as usual $\otimes$ denotes the tensor product. Introduce generic vectors 
\begin{eqnarray}
\label{eq:tensordef}
{\bf r} & := & \sum_{\substack{a,b \in [n] \\ a \ne b \\ \ell \in [N]}} r_{ab}(\ell) \cdot {\bf e}^{(n)}_a \otimes {\bf e}^{(n)}_b \otimes {\bf e}^{(N)}_\ell; \nonumber \\
{\bf p} & := & \sum_{\ell, \ell' \in [N]} p_{s(\ell)t(\ell)}(\ell') \cdot {\bf e}^{(N)}_\ell \otimes {\bf e}^{(N)}_{\ell'}; \nonumber \\
{\bf \pi} & := & \sum_{\ell, \ell' \in [N]} \pi_\ell(\ell') \cdot {\bf e}^{(N)}_\ell \otimes {\bf e}^{(N)}_{\ell'}; \nonumber \\
{\bf \lambda} & := & \sum_{\ell \in [N]} \lambda_\ell {\bf e}^{(N)}_\ell
\end{eqnarray}
and a scalar $\mu$ corresponding to the (actual plus virtual) \emph{maximum link utilization} as building blocks for
\begin{equation}
\label{eq:compositevector}
{\bf x} := {\bf r} \oplus {\bf p} \oplus {\bf \pi} \oplus {\bf \lambda} \oplus \mu.
\end{equation}
Here we recall that direct sum of ${\bf v}$ and ${\bf w}$ is ${\bf v} \oplus {\bf w} := ({\bf v}^T, {\bf w}^T)^T$, so that ${\bf x} \in (\mathbb{R}^n \otimes \mathbb{R}^n \otimes \mathbb{R}^N) \oplus (\mathbb{R}^N \otimes \mathbb{R}^N) \oplus (\mathbb{R}^N \otimes \mathbb{R}^N) \oplus \mathbb{R}^N \oplus \mathbb{R} \cong \mathbb{R}^{n^2 N + N^2 + N^2 + N + 1}$. 

As a preliminary step \emph{en route} to obtaining an equivalent of \eqref{eq:R3LPprecursor} in terms of ${\bf x}$, we shall first express \eqref{eq:routing} in explicit matrix form. Towards this end, first note that 
\begin{equation}
\label{eq:routingtensor}
\sum_{\ell : s(\ell) = j} r_{ab}(\ell) = \left ( \sum_{\ell : s(\ell) = j} {\bf e}^{(n)}_a \otimes {\bf e}^{(n)}_b \otimes {\bf e}^{(N)}_\ell \right ) \cdot {\bf r},
\end{equation}
immediately gives an implicit matrix form for \eqref{eq:RoutingSelf}-\eqref{eq:RoutingNoExtension} that is readily made explicit \emph{in silico} as
\begin{equation}
\label{eq:Rr}
R{\bf r} = {\bf \rho}.
\end{equation}
A similar equation 
\begin{equation}
\label{eq:Pp}
P{\bf p} = {\bf \rho} \odot \sigma
\end{equation} 
encodes the requirement that $p$ be a routing: here $\odot$ denotes the entrywise product. Let \texttt{ind\_R} be an array formed by stacking rows $(a,b,\ell)$ in lexicographic order. The following MATLAB snippet indicates how to obtain $P$ and $\sigma$:

{\footnotesize
\begin{verbatim}
% L is a Nx2 array of link sources and targets
P = [];
ind_P = [];
sigma = zeros(size(R,1),1);
for ell = 1:size(L,1)
    ind = ismember(ind_R(:,1:2),L(ell,:),'rows');
    P = [P,R(:,ind)];
    ind_P = [ind_P;ind_R(ind,:)];
    sigma = sigma+any(R(:,ind),2);
end
\end{verbatim}
}

The specification of $R$ (up to signs of rows that are irrelevant and may be chosen freely) and $\rho$ can be completed by proceeding through the scalar equations of \eqref{eq:routing} in order and subsequently eliminating trivial or redundant equations in the order they are encountered, so that $N_R \le {nN + n(n-1)(n-2) + 2n(n-1) + 2(n-1)N}$ scalar equations remain, i.e. $R$ is a $N_R \times n^2 N$ matrix and $\rho$ is a vector of dimension $N_R$. The bound on $N_R$ arises as follows: \eqref{eq:RoutingSelf} gives $nN$ scalar equations; \eqref{eq:RoutingConservation} gives $n(n-1)(n-2)$ scalar equations (ignoring the possibility of sources/targets); \eqref{eq:RoutingOutTotality} and \eqref{eq:RoutingInTotality} each give $n(n-1)$ scalar equations, and \eqref{eq:RoutingNoReturn} and \eqref{eq:RoutingNoExtension} each give $(n-1)N$ scalar equations.

\section{\label{sec:R3LP}R3 linear program}

The term $\max_{z \in Z_F} \sum_{\ell'} z_{\ell'} p_{s(\ell')t(\ell')}(\ell)$ in \eqref{eq:R3LPprecursor} is the optimal objective of $\max_{\bf \zeta} \langle {\bf \chi}, {\bf \zeta} \rangle$ subject to $A {\bf \zeta} \le {\bf \beta}$ and ${\bf \zeta} \ge {\bf 0}$, where $\chi_{\ell'} := c_{\ell'} p_{s(\ell')t(\ell')}(\ell)$, $\zeta_\ell := z_\ell/c_\ell$, $A := \left ( \begin{smallmatrix} {\bf 1}^T \\ I \end{smallmatrix} \right )$, and $\beta := \left ( \begin{smallmatrix} F \\ {\bf 1} \end{smallmatrix} \right )$. This optimal objective is the same as that of the dual linear program $\min_\omega \langle \beta, \omega \rangle$ subject to $A^T {\bf \omega} \ge {\bf \chi}$ and ${\bf \omega} \ge {\bf 0}$. 

Writing $\omega =: (\lambda_\ell, \pi_\ell(1), \dots, \pi_\ell(N))^T$, this dual linear program is (after some trivial rearrangements)
\begin{subequations}
\begin{align}
\min_{\lambda_\ell, \pi_\ell(\cdot)} \left ( \sum_{\ell'} \pi_\ell(\ell') + \lambda_\ell F \right ) \quad \text{s.t.}
	& \nonumber \\
c_{\ell'} p_{s(\ell')t(\ell')}(\ell) - \pi_\ell(\ell') - \lambda_\ell \quad \le 
	& \quad 0; \nonumber \\
\pi_\ell(\ell'), \lambda_\ell \quad \ge 
	& \quad 0. \nonumber 
\end{align}
\end{subequations}
From here we immediately get the \emph{R3LP} linear program (for $F$ a fixed positive integer)
\begin{subequations}
\label{eq:R3LP}
\begin{align}
\min_{{\bf x}} \mu \quad \text{s.t. $r$ and $p$ are routings};
	& \\
\sum_{a,b} d_{ab} r_{ab}(\ell) + \sum_{\ell'} \pi_\ell(\ell') + \lambda_\ell F - c_\ell \mu \quad \le 
	& \quad 0; \label{eq:R3LPDual1} \\
c_{\ell'} p_{s(\ell') t(\ell')}(\ell) - \pi_\ell(\ell') - \lambda_\ell \quad \le 
	& \quad 0; \label{eq:R3LPDual2} \\
\pi_\ell(\ell'), \lambda_\ell \quad \ge 
	& \quad 0. \label{eq:R3LPLowerUpper}
\end{align}
\end{subequations}
Note that \eqref{eq:R3LP} has obvious variants called in which semiroutings and restricted semiroutings are considered instead. 

The remaining details are as follows. Let ${\bf 0}_m$, ${\bf 1}_m$, and ${\bf \infty}_m$ denote the column vectors with $m$ entries all equal to $0$, $1$, or $\infty$, respectively; we may also write, e.g., ${\bf 0}_m \equiv 0_{m \times 1}$, where $0_{m \times m'}$ is a $m \times m'$ matrix with all entries equal to zero. Define the block matrices
\begin{equation}
A^{(=)} := \left ( \begin{smallmatrix}
R & 0_{N_R \times N^2} & 0_{N_R \times N^2} & 0_{N_R \times N} & 0_{N_R \times 1} \\
0_{N_R \times n^2 N} & P & 0_{N_R \times N^2} & 0_{N_R \times N} & 0_{N_R \times 1}
\end{smallmatrix} \right )
\end{equation}
and
\begin{equation}
A^{(\le)} := \left ( \begin{smallmatrix}
{\bf d}^\flat \otimes I_N & 0_{N \times N^2} & I_N \otimes {\bf 1}_N^T & F I_N & -{\bf c}^\sharp \\
0_{N^2 \times n^2 N} & \Delta({\bf c}^\sharp) \otimes I_N & -S & -I_N \otimes {\bf 1}_N & 0_{N^2 \times 1} 
\end{smallmatrix} \right )
\end{equation}
where ${\bf d}^\flat := (d_{11},d_{12},\dots,d_{nn})$, $I_N$ is the $N$-dimensional identity matrix, ${\bf c}^\sharp := (c_1,c_2,\dots,c_N)^T$, $\Delta$ denotes the diagonal operation, and $S$ is an involutory permutation matrix of dimension $N^2$ that effectively swaps link indices \emph{\`a la} $(\ell,\ell') \leftrightarrow (\ell',\ell)$ and that is conveniently defined as follows: 
\begin{equation}
S_{jk} := \begin{cases}
1 & \text{ if } k = N \cdot ((j-1) \text{ mod } N) + 1 + \left \lfloor \frac{j-1}{N} \right \rfloor \\
0 & \text{ otherwise.}
\end{cases}
\end{equation}
Writing ${\bf b}^{(=)} := {\bf \rho} \oplus ({\bf \rho} \odot \sigma)$ and ${\bf u} := {\bf 1}_{n^2 N} \oplus {\bf 1}_{N^2} \oplus {\bf \infty}_{N^2} \oplus {\bf \infty}_N \oplus \infty$, R3LP takes the MATLAB-ready form
\begin{equation}
\label{eq:R3LPmatrix}
\min_{{\bf x}} \mu \text{ s.t. } A^{(=)}{\bf x} = {\bf b}^{(=)}; \ A^{(\le)}{\bf x} \le {\bf 0}; \ {\bf x} \ge {\bf 0}; \ {\bf x} \le {\bf u}.
\end{equation}

\section{\label{sec:Technicalities}Technicalities}

\subsection{\label{sec:Parallel}Dealing with parallel links}

There is no problem with defining $r$ when there are parallel links. However, there is a serious but subtle problem with defining $p$ that is manifested by components of ${\bf p}$ that are structurally forced to be equal. Note that the source/target pairs $(s(\ell),t(\ell))$ are distinct iff there are no parallel links. In this case \emph{only} we can regard $L$ as a subset of $V^2$. In the event that there are parallel links, the notion of a ``protection routing'' as embodied by $p$ becomes either ill-defined (unless all parallel links have the same capacity) or useless (since parallel links need not have the same capacity).

That is, we must regard $p$ as a function on $L \times L$ or on $V^2 \times L$. Both cases can apply if there are no parallel links, since then there is a bijection between $L$ and the set of unique source/target pairs $U := \{(s(\ell),t(\ell)) : \ell \in [N]\} \subseteq V^2$, and we can regard $p$ as a function on $V^2 \times L$ which is zero outside of $U \times L$. But if there are parallel links and only the first case applies, then the expression $p_{s(\ell)t(\ell)}(\ell')$ cannot be assigned a consistent meaning unless it takes the same value for all parallel links $\ell$. But this is essentially the second case, and then the notion of the protection routing generally becomes useless, since there is then no way to completely account for parallel links with different capacities. The inextricability of the protection routing and link capacities is also latent in the matrix formulation of \S \ref{sec:R3LP}, which turns out to rest in an essential way on interpreting $p$ as a function on $L \times L$.

In trying to cut this Gordian knot, the obvious tactic is to insert virtual vertices and links. However, this introduces new problems. For instance, suppose that every parallel link is split into two links joined at a virtual vertex. Then while this eliminates any internal inconsistency associated with $p$, it also introduces a degeneracy into R3LP that forces $\mu > 1$, obliterating the non-congestion guarantee for $\mu < 1$ that is at the heart of R3. Furthermore, experiments (not detailed here) show that removing constraints \eqref{eq:R3LPDual1} and \eqref{eq:R3LPDual2} associated with either the ``outgoing half'' or ``incoming half'' of the new links does not fix this problem, which turns out to be due to entries of the form $p_\ell(\ell)$ that can be ``carefully ignored.''

It seems unlikely that more elaborate virtual topology schemes (e.g., splitting vertices) would succeed where the one sketched above fail. In any event, we have searched for but have not found such a scheme that works. Additionally, while it is conceivable that simultaneously fusing parallel links and altering the rerouting virtual demand set in \cite{WangEtAl} could be done in such a way as to address the case of $F \le 1$ failures, it seems unlikely that such a strategy could ever work for $F > 1$.

The underlying degeneracy that is introduced by topology virtualization turns out to be protection routing values of the form $p_\ell(\ell) = 1$. As \cite{WangEtAl} points out, for $\mu \le 1$ on the original topology, an equality $p_\ell(\ell) = 1$ 
\begin{quote}
``implies that link [$\ell$] carries no actual demand from [source-target] pairs or virtual demand from links other than [$\ell$]. So link [$\ell$] does not need to be protected and can be safely ignored.''
\end{quote}
With this in mind, we can evaluate the maximum load, given as the optimal objective to 
\begin{equation}
\label{eq:MaximumLoad}
\max_z \sum_\ell z_\ell p_\ell(\ell') \text{ s.t. } z_\ell \le c_\ell \text{ and } \sum_\ell z_\ell/c_\ell \le F, \nonumber
\end{equation}
with protection routing values of the form $p_\ell(\ell) = 1$ either left unchanged or reset to zero, and compare these results with the dual objective $\sum_\ell \pi_{\ell'}(\ell) + \lambda_{\ell'} F$. 

In practice, values $\mu > 1$ include constributions from ignorable diagonal protection routing values, and properly accounting for such cases after a topology virtualization allows us to recapture guarantees of congestion-free routing.

\subsection{\label{sec:RoutingPreservation}Preservation of routing constraints}

It turns out that the reconfiguration scheme of \cite{WangEtAl} does not actually enforce \eqref{eq:routing}. It is clear that \eqref{eq:RoutingSelf} continues to hold and easy to show (using the fact that the original base and protection routings satisfy \eqref{eq:RoutingConservation}) that \eqref{eq:RoutingConservation} also continues to hold. But \eqref{eq:RoutingOutTotality}, \eqref{eq:RoutingInTotality}, \eqref{eq:RoutingNoReturn}, and \eqref{eq:RoutingNoExtension} do not automatically continue to hold. In fact, it is not hard to construct an example in which traffic is routed along a cycle after reconfiguration.

Though this problem is irksome, it is not critical: auxiliary techniques (e.g., forwarding only once, flow decomposition, or prohibiting turns \cite{LevitinKarpovskyMustafa}) can ameliorate it, and like the reconfiguration as a whole, it is a transient issue that lasts only until a new base routing can be solved for. It is also plausible that additional constraints along the lines of $r_{ab}(\ell) = r_{a'b}(\ell)$ might circumvent the problem altogether.

\section{\label{sec:WirelessR3}Wireless R3}

\subsection{\label{sec:WirelessFormalism}Formalism}

A formalism for wireless networks requires the capability to describe point-to-multipoint (P2MP) transmission. 
\footnote{
Multipoint-to-point reception can be described similarly.
}
Towards this end, we introduce some notation before giving a toy example. Let $L_j := \{\ell : s(\ell) = j\}$ be the set of links with source vertex $j$. For $\Gamma_j \in \mathbb{N}$, let $\gamma_j : L_j \rightarrow [\Gamma_j]$ be a surjective function: for each $g \in \Gamma_j$, the preimage $\gamma_j^{-1}(g)$ is the set of links belonging to the $g$th \emph{P2MP group} at vertex $j$. A singleton group corresponds to a dedicated point-to-point transmission. 

Noting that $L = \cup_j L_j$ and writing $\Gamma := \cup_j (\{j\} \times [\Gamma_j])$, we can summarize the additional structure for P2MP transmission in the commutative diagram (i.e., a digraph with edges labeled by functions such that function compositions corresponding to paths with the same source and target give the same results)
\begin{equation}
\begin{tikzpicture}[->,>=stealth',shorten >=1pt]
	\node (E) at (0,0) {$L$};
	\node (Ej) at (0,1.5) {$L_j$};
	\node (Gj) at (2,1.5) {$[\Gamma_j]$};
	\node (G) at (2,0) {$\Gamma$};
	\node (R) at (4,.75) {$\mathbb{R}$};
	\path [right hook->] (Ej) edge node [left] {$i$} (E);
	\path [right hook->] (Gj) edge node [left] {$i$} (G);
	\path [->] (Ej) edge node [above] {$\gamma_j$} (Gj);
	\path [->] (E) edge node [below] {$\gamma$} (G);
	\path [->] (Gj) edge node [above] {$c_j$} (R);
	\path [->] (G) edge node [below] {$c$} (R);
\end{tikzpicture} \nonumber
\end{equation}
where here $i$ indicates a generic inclusion. The \emph{group capacity} is given in terms of a family of vertex-specific maps $c_j$ via $c(j,g) := c_j(g)$ and for $\ell \in L_j$ we have $\gamma(\ell) := (j,\gamma_j(\ell))$.

\subsection{\label{sec:WirelessFormalismExample}Example}

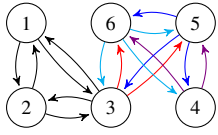
\begin{figure}
	\begin{center}
	\begin{tikzpicture}[scale=0.75,->,>=stealth',shorten >=1pt,every node/.style={transform shape}]
		\node [draw,circle,minimum size=7mm] (v1) at (0,1.5) {1};
		\node [draw,circle,minimum size=7mm] (v2) at (0,0) {2};
		\node [draw,circle,minimum size=7mm] (v3) at (1.5,0) {3};
		\node [draw,circle,minimum size=7mm] (v4) at (3,0) {4};
		\node [draw,circle,minimum size=7mm] (v5) at (3,1.5) {5};
		\node [draw,circle,minimum size=7mm] (v6) at (1.5,1.5) {6};
		\draw (v1) [out=-108,in=108,looseness=1] to (v2);
		\draw (v1) [out=-54,in=144,looseness=1] to (v3);
		\draw (v2) [out=72,in=-72,looseness=1] to (v1);
		\draw (v2) [out=-18,in=-162,looseness=1] to (v3);
		\draw (v3) [out=126,in=-36,looseness=1] to (v1);
		\draw (v3) [out=162,in=18,looseness=1] to (v2);
		\draw (v3) [color=red,out=36,in=-126,looseness=1] to (v5);
		\draw (v3) [color=red,out=72,in=-72,looseness=1] to (v6);
		\draw (v4) [color=violet,out=72,in=-72,looseness=1] to (v5);
		\draw (v4) [color=violet,out=126,in=-36,looseness=1] to (v6);
		\draw (v5) [color=blue,out=-144,in=54,looseness=1] to (v3);
		\draw (v5) [color=blue,out=-108,in=108,looseness=1] to (v4);
		\draw (v5) [color=blue,out=162,in=18,looseness=1] to (v6);
		\draw (v6) [color=cyan,out=-108,in=108,looseness=1] to (v3);
		\draw (v6) [color=cyan,out=-54,in=144,looseness=1] to (v4);
		\draw (v6) [color=cyan,out=-18,in=-162,looseness=1] to (v5);
	\end{tikzpicture}
	\end{center}
	\caption{\label{fig:WirelessExampleTopology} $G$: $n = 6$ and $N = 16$. Vertices 1-3 represent fixed terrestrial sites, with vertex 3 a coastal communications station; vertex 4 represents a ship (over the horizon from the station), vertex 5 represents a plane (within range of the station and ship), and vertex 6 represents an overhead satellite. Nontrivial P2MP groups are shown in {\color{red}red}, {\color{violet}violet}, {\color{blue}blue}, and {\color{cyan}cyan}.}
\end{figure}

We illustrate \S \ref{sec:WirelessFormalism} with an example. Figure \ref{fig:WirelessExampleTopology} depicts the underlying digraph $G$ and P2MP groups of a network in which the communications between three fixed terrestrial nodes, a ship, a plane, and a satellite are cariacatured. 

By inspection, we have $\Gamma_1 = 2 = |L_1|$, $\Gamma_2 = 2 = |L_2|$, $\Gamma_3 = 3 < |L_3| = 4$, $\Gamma_4 = 1 < |L_4| = 2$, $\Gamma_5 = 1 < |L_5| = 3$, and $\Gamma_6 = 1 < |L_6| = 3$. Assuming (by default) that in the absence of parallel links the link indices correspond to the lexicographical ordering of source/target pairs, the maps $\gamma_j$ are given (without loss of generality) by $\gamma_1^{\times 2}(1,2) = (1,2)$; $\gamma_2^{\times 2}(3,4) = (1,2)$; $\gamma_3^{\times 4}(5,6,{\color{red}7},{\color{red}8}) = (1,2,{\color{red}3},{\color{red}3})$; $\gamma_4^{\times 2}({\color{violet}9},{\color{violet}10}) = ({\color{violet}1},{\color{violet}1})$; $\gamma_5^{\times 3}({\color{blue}11},{\color{blue}12},{\color{blue}13}) = ({\color{blue}1},{\color{blue}1},{\color{blue}1})$, and $\gamma_6^{\times 3}({\color{cyan}14},{\color{cyan}15},{\color{cyan}16}) = ({\color{cyan}1},{\color{cyan}1},{\color{cyan}1})$.
%\begin{eqnarray}
%\gamma_1^{\times 2}(1,2) & = & (1,2); \nonumber \\
%\gamma_2^{\times 2}(3,4) & = & (1,2); \nonumber \\
%\gamma_3^{\times 4}(5,6,{\color{red}7},{\color{red}8}) & = & (1,2,{\color{red}3},{\color{red}3}); \nonumber \\
%\gamma_4^{\times 2}({\color{violet}9},{\color{violet}10}) & = & ({\color{violet}1},{\color{violet}1}); \nonumber \\
%\gamma_5^{\times 3}({\color{blue}11},{\color{blue}12},{\color{blue}13}) & = & ({\color{blue}1},{\color{blue}1},{\color{blue}1}); \nonumber \\
%\gamma_6^{\times 3}({\color{cyan}14},{\color{cyan}15},{\color{cyan}16}) & = & ({\color{cyan}1},{\color{cyan}1},{\color{cyan}1}). \nonumber
%\end{eqnarray}
The lexicographic ordering on links carries over to elements of $\Gamma$, and $c^{\times N}(1,\dots,N) = (c_1(1), c_1(2), c_2(1), c_2(2), c_3(1), c_3(2), {\color{red}c_3(3)}, {\color{red}c_3(3)}, {\color{violet}c_4(1)}, \\ {\color{violet}c_4(1)}, {\color{blue}c_5(1)}, {\color{blue}c_5(1)}, {\color{blue}c_5(1)}, {\color{cyan}c_6(1)}, {\color{cyan}c_6(1)}, {\color{cyan}c_6(1)})$. $\Box$

\subsection{\label{sec:WirelessFormalismConstraint}Wireless constraint}

Absent parallel links, the additional constraint imposed by wireless communications can now be written down: 
\begin{equation}
\label{eq:WirelessConstraint}
\sum_{\ell \in \gamma_j^{-1}(g)} d_{j t(\ell)} r_{j t(\ell)} (\ell) \le c_j(g),
\end{equation}
where $g \in [\Gamma_j]$ and $|\gamma_j^{-1}(g)| > 1$ to avoid redundancy.

\subsection{\label{sec:WirelessFormalismExample2}Example 2}

The presence of parallel links introduces additional intricacy which we illustrate through an example. Consider $G$ as in the left panel of Figure \ref{fig:WirelessExampleTopology2}. 
\begin{figure}
	\begin{center}
	\begin{tikzpicture}[scale=.5,->,>=stealth',shorten >=1pt,every node/.style={transform shape}]
		\node [draw,circle,minimum size=7mm] (v1) at (0,0) {1};
		\node [draw,circle,minimum size=7mm] (v2) at (4,0) {2};
		\node [draw,circle,minimum size=7mm] (v3) at (2.3511,3.2361) {3};	% 54 deg
		\node [draw,circle,minimum size=7mm] (v4) at (0.0000,4.0000) {4};	
		\node [draw,circle,minimum size=7mm] (v5) at (-3.2361,2.3511) {5};	% 144 deg
		\node [draw,circle,minimum size=7mm] (v6) at (-3.2361,-2.3511) {6};	% -144 deg
		\node [draw,circle,minimum size=7mm] (v7) at (-0.0000,-4.0000) {7};
		\node [draw,circle,minimum size=7mm] (v8) at (2.3511,-3.2361) {8};	% -54 deg
		\path [->,color=red,out=-18,in=-162,looseness=1] (v1) edge node [below] {{\color{black}1}} (v2);
		\path [->,color=violet,out=18,in=162,looseness=1] (v1) edge node [above] {{\color{black}2}} (v2);
		\path [->,color=red] (v1) edge node [right] {{\color{black}3}} (v3);
		\path [->,color=violet] (v1) edge node [right] {{\color{black}4}} (v4);
		\path [->,color=red,out=126,in=-18,looseness=1] (v1) edge node [above] {{\color{black}5}} (v5);
		\path [->,color=blue,out=162,in=-54,looseness=1] (v1) edge node [below] {{\color{black}6}} (v5);
		\path [->,color=violet,out=-162,in=54,looseness=1] (v1) edge node [above] {{\color{black}7}} (v6);
		\path [->,color=blue,out=-126,in=18,looseness=1] (v1) edge node [below] {{\color{black}8}} (v6);
		\path [->,color=blue] (v1) edge node [left] {{\color{black}9}} (v7);
		\path [->,color=cyan] (v1) edge node [below] {{\color{black}10}} (v8);
	\end{tikzpicture} 
	\quad 
	\begin{tikzpicture}[scale = .5,->,>=stealth',shorten >=1pt,every node/.style={transform shape}]
		\node [draw,circle,minimum size=7mm] (v1) at (0,0) {1};
		\node [draw,circle,minimum size=7mm] (v2) at (4,0) {2};
		\node [draw,circle,minimum size=7mm] (v3) at (2.3511,3.2361) {3};	% 54 deg
		\node [draw,circle,minimum size=7mm] (v4) at (0.0000,4.0000) {4};	
		\node [draw,circle,minimum size=7mm] (v5) at (-3.2361,2.3511) {5};	% 144 deg
		\node [draw,circle,minimum size=7mm] (v6) at (-3.2361,-2.3511) {6};	% -144 deg
		\node [draw,circle,minimum size=7mm] (v7) at (-0.0000,-4.0000) {7};
		\node [draw,circle,minimum size=7mm] (v8) at (2.3511,-3.2361) {8};	% -54 deg
		\node [draw,circle,minimum size=7mm] (v9) at (1.9021,-0.6180) {9};	% -18 deg
		\node [draw,circle,minimum size=7mm] (v10) at (1.9021,0.6180) {10};	% 18 deg
		\node [draw,circle,minimum size=7mm] (v11) at (-1.1756,1.6180) {11};	% 126 deg
		\node [draw,circle,minimum size=7mm] (v12) at (-1.9021,0.6180) {12};	% 162 deg
		\node [draw,circle,minimum size=7mm] (v13) at (-1.9021,-0.6180) {13};	% -162 deg
		\node [draw,circle,minimum size=7mm] (v14) at (-1.1756,-1.6180) {14};	% -126 deg
		\draw (v1) [color=red] to (v9);		
		\draw (v1) [color=violet] to (v10);		
		\draw (v1) [color=red] to (v3);		
		\draw (v1) [color=violet] to (v4);		
		\draw (v1) [color=red] to (v11);		
		\draw (v1) [color=blue] to (v12);		
		\draw (v1) [color=violet] to (v13);		
		\draw (v1) [color=blue] to (v14);		
		\draw (v1) [color=blue] to (v7);		
		\draw (v1) [color=cyan] to (v8);		
		\foreach \from/\to in {
			v9/v2, v10/v2, v11/v5, v12/v5, v13/v6, v14/v6}
			\draw (\from) to (\to);
	\end{tikzpicture} 
	\end{center}
	\caption{\label{fig:WirelessExampleTopology2} (L) $G$: $n = 8$ and $N = 10$. Links are ordered counterclockwise starting at the red link from vertex 1 to vertex 2 (note that this is one of $2^3$ possible lexicographic orderings). We have that $\Gamma_1 = 4$ and $\gamma_1^{\times 10}(1,\dots,10) = ({\color{red}1},{\color{violet}2},{\color{red}1},{\color{violet}2},{\color{red}1},{\color{blue}3},{\color{violet}2},{\color{blue}3},{\color{blue}3},{\color{cyan}4})$. (R) After virtualization, there are no parallel links, so we order the links lexicographically by source/target pair. This effectively reorders the links in $L_1$, so that now $\gamma_1^{\times 10}(1,\dots,10) = ({\color{red}1},{\color{violet}2},{\color{blue}3},{\color{cyan}4},{\color{red}1},{\color{violet}2},{\color{red}1},{\color{blue}3},{\color{violet}2},{\color{blue}3})$.}
\end{figure}
We have $\gamma_1^{-1}({\color{red}1}) = \{1,3,5\}$, $\gamma_1^{-1}({\color{violet}2}) = \{2,4,7\}$, and $\gamma_1^{-1}({\color{blue}3}) = \{6,8,9\}$. These yield the following instances of \eqref{eq:WirelessConstraint}:
\begin{eqnarray}
d_{12} r_{12}(1) + d_{13} r_{13}(3) + d_{15} r_{15}(5) & \le & c_1({\color{red}1}); \nonumber \\ 
d_{12} r_{12}(2) + d_{14} r_{14}(4) + d_{16} r_{16}(7) & \le & c_1({\color{violet}2}); \nonumber \\ 
d_{15} r_{15}(6) + d_{16} r_{16}(8) + d_{17} r_{17}(9) & \le & c_1({\color{blue}3}); \nonumber \\
d_{18} r_{18}(10) & \le & c_1({\color{cyan}4}). \nonumber
\end{eqnarray}
Virtualization adds vertices after $n = 8$, with the (now unique) convention that links are ordered lexicographically. Thus, referencing the right panel of Figure \ref{fig:WirelessExampleTopology2}, the preceding constraints correspond after virtualization to 
\begin{eqnarray}
d_{12} r_{12}(5) + d_{13} r_{13}(1) + d_{15} r_{15}(7) & \le & c_1({\color{red}1}); \nonumber \\ 
d_{12} r_{12}(6) + d_{14} r_{14}(2) + d_{16} r_{16}(9) & \le & c_1({\color{violet}2}); \nonumber \\ 
d_{15} r_{15}(8) + d_{16} r_{16}(10) + d_{17} r_{17}(3) & \le & c_1({\color{blue}3}); \nonumber \\
d_{18} r_{18}(4) & \le & c_1({\color{cyan}4}). \nonumber
\end{eqnarray}

\subsection{\label{sec:WirelessFormalismConstraintGeneralized}Generalized wireless constraint}

The example of \S \ref{sec:WirelessFormalismExample2} above illustrates that if we use $\sigma^{-1}$ to denote the permutation relating the original and virtualized links (e.g., for the case of Figure \ref{fig:WirelessExampleTopology2}, $\sigma^{-1} = (3,4,9,10,1,2,5,6,7,8)$ and $\sigma = (5,6,1,2,7,8,9,10,3,4)$), then the generalization of \eqref{eq:WirelessConstraint} to incorporate parallel links is (again with $g \in [\Gamma_j]$ and $|\gamma_j^{-1}(g)| > 1$)
\begin{equation}
\label{eq:WirelessConstraintParallel}
\sum_{\ell \in \gamma_j^{-1}(g)} d_{j t(\ell)} r_{j t(\ell)} (\sigma(\ell)) \le c_j(g).
\end{equation}
Care must be taken in the interpretation of each side of \eqref{eq:WirelessConstraintParallel}: the vertex and link indices refer to the original topology, but the indexed objects themselves (in particular, the routing $r$) are defined for the virtualized topology. Note that if there are no parallel links in the original topology, then $\sigma$ is the identity permutation, so \eqref{eq:WirelessConstraintParallel} does in fact generalize \eqref{eq:WirelessConstraint}.

\section{\label{sec:WR3LP}A realistic example}

\begin{figure*}[htbp]
\centering{
\includegraphics[trim = 0mm 0mm 0mm 0mm, clip, width=\textwidth,keepaspectratio]{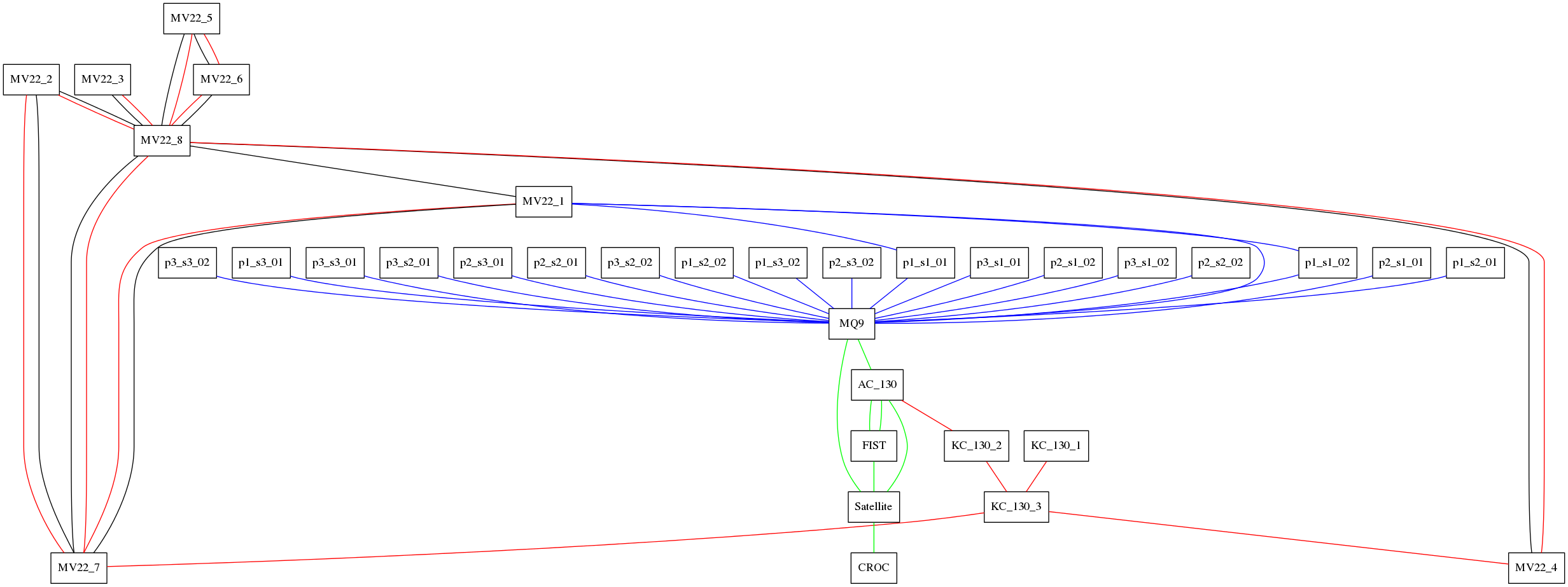}% Here is how to import pix
}
\caption{ \label{fig:TacticalExampleWireless}A network with link capacities color-coded as follows: black links have capacity $10^{-4}$ Gbps; {\color{red}red links, $10^{-3}$ Gbps}; {\color{blue}blue links, $10^{-2}$ Gbps}; and {\color{green}green links, $10^{-1}$ Gbps}. P2MP groups are defined by color: i.e., each set of links from a given vertex with a given color defines a P2MP group.} 
\end{figure*}

\eqref{eq:R3LP} and \eqref{eq:WirelessConstraintParallel} yield a linear program that we apply to a realistic example. Given a capacity function $c$ such as that in Figure \ref{fig:TacticalExampleWireless}, we define a simple but reasonable toy model of traffic demand as follows. First, note that we can define a probability distribution $\mathbb{P}$ on $V$ using the well-known PageRank for weighted directed multigraphs \cite{PageEtAl}. 
\footnote{
NB. We use a PageRank ``damping factor'' of 0.85.
} 
Second, we choose some $D \ll 1$ and define our model demand $d_{ab} := D \cdot (\sum_{\ell : s(\ell) = a} c_\ell) \cdot \mathbb{P}(b)$ via
\begin{equation}
\label{eq:DemandModel}
\frac{d_{ab}}{\sum_{b'} d_{ab'}} = \mathbb{P}(b); \quad \sum_b d_{ab} = D \sum_{\ell : s(\ell) = a} c_\ell.
\end{equation}
Note that $P_{ab} := \frac{d_{ab}}{\sum_{b'} d_{ab'}}$ are the entries of a row-stochastic matrix. Thus the first equation in \eqref{eq:DemandModel} embodies the intuition that $\mathbb{P}$ is an invariant measure for the Markov chain defined by $P$, which is broadly consistent with the idea behind PageRank. Meanwhile, the second equation in \eqref{eq:DemandModel} above merely says that outbound demand is proportional to outbound capacity. 

\begin{figure}[htbp]
\centering{
\includegraphics[trim = 0mm 5mm 5mm 5mm, clip, width=\columnwidth,keepaspectratio]{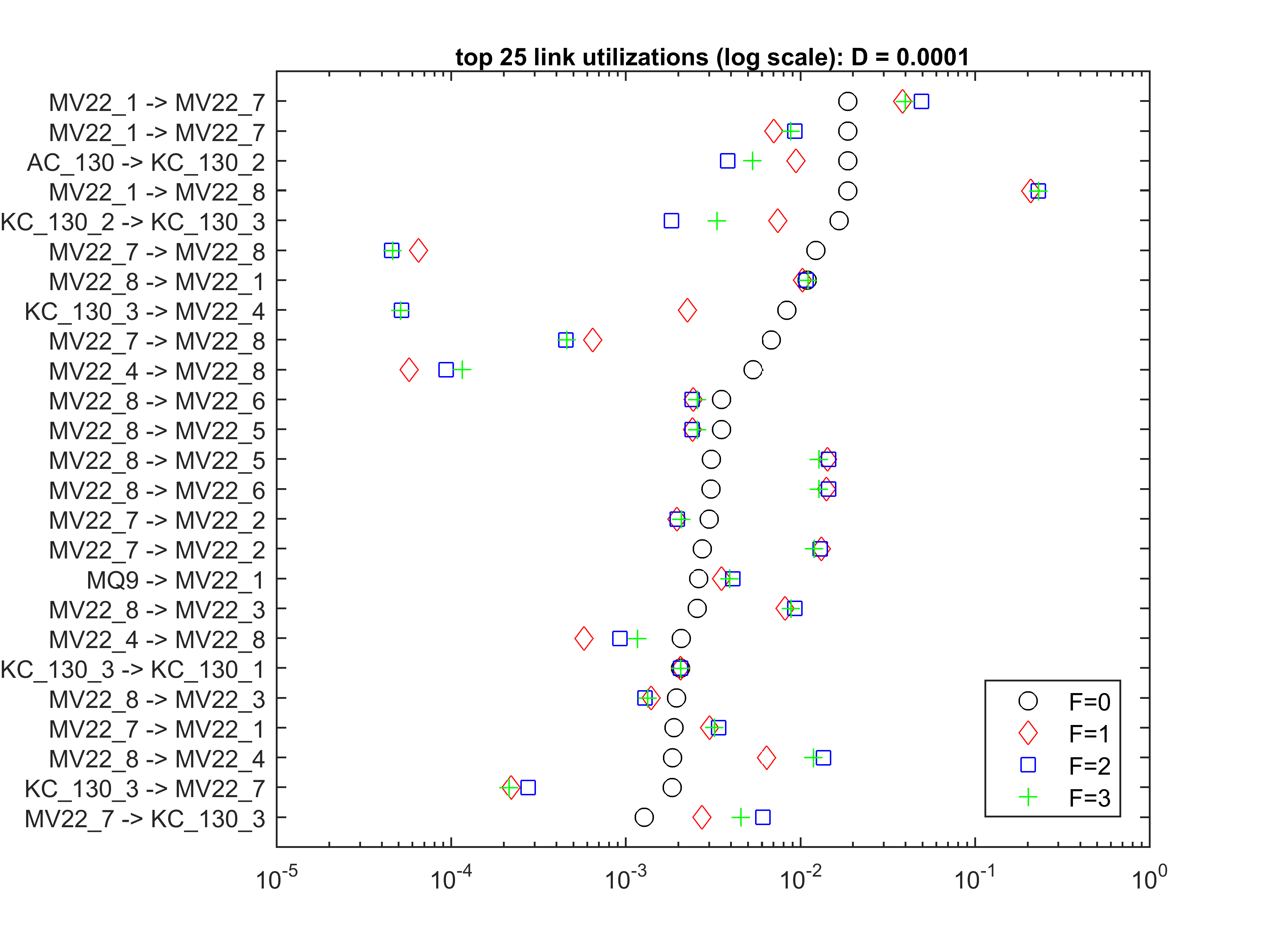}% Here is how to import pix
}
\caption{ \label{fig:TopLinkUtilizationsLow} Top link utilizations for low ($D = 10^{-4}$) demand and varying $F$. } 
\end{figure}

\begin{figure}[htbp]
\centering{
\includegraphics[trim = 0mm 5mm 5mm 5mm, clip, width=\columnwidth,keepaspectratio]{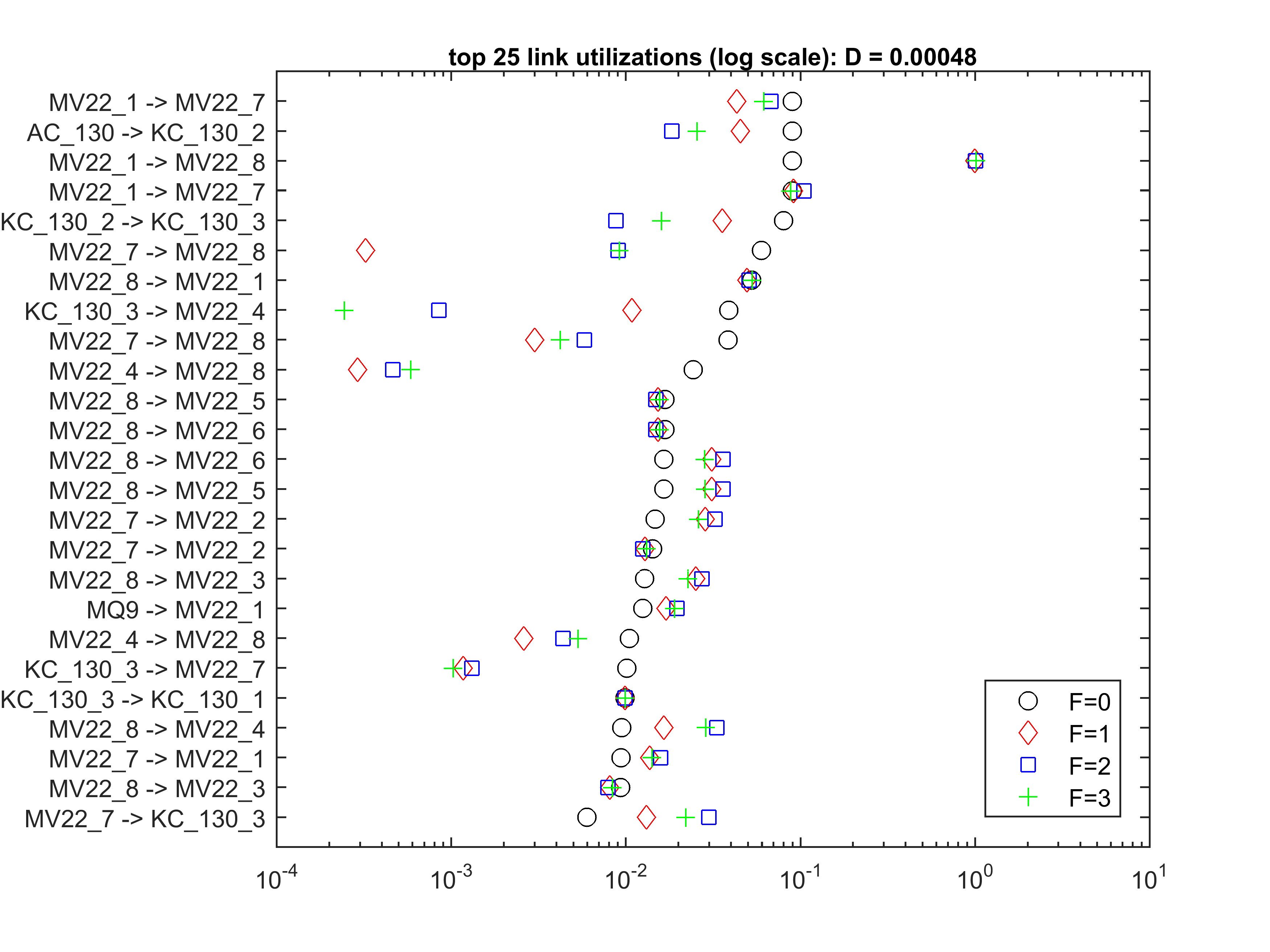}% Here is how to import pix
}
\caption{ \label{fig:TopLinkUtilizationsHigh} As in Figure \ref{fig:TopLinkUtilizationsLow} with high ($D = 4.8 \cdot 10^{-4}$) demand. } 
\end{figure}

As Figures \ref{fig:TopLinkUtilizationsLow} and \ref{fig:TopLinkUtilizationsHigh} show, the directed link from \texttt{MV22\_1} to \texttt{MV22\_8} becomes overwhelmed for $D \approx 4.8 \cdot 10^{-4}$, at which point all other links (including from \texttt{MV22\_8} to \texttt{MV22\_1}) have utilization below about $0.1$. Note that the links between \texttt{MV22\_1} to \texttt{MV22\_8} form a obvious bottleneck, illustrating how our approach can be used for planning purposes.

\section*{Acknowledgements}
This work was supported by the Office of Naval Research under contract N00014-15-C-5102. The authors thank Jeong-O Jeong for his help in applying the techniques here in a detailed simulation environment.


\vspace*{-0.25cm}

\begin{thebibliography}{10}

\bibitem{Chiesa1}Chiesa, M. \emph{et al.} ``On the resiliency of randomized routing against multiple edge failures.'' ICALP (2016).

\bibitem{Chiesa2}Chiesa, M. \emph{et al.} ``The quest for resilient (static) forwarding tables.'' INFOCOM (2016).

\bibitem{Chiesa3}Chiesa, M. \emph{et al.} ``On the resiliency of static forwarding tables.'' \emph{IEEE Trans. Networking} {\bf 25}, 1133 (2017).

\bibitem{Foerster}Foerster, K.-T. \emph{et al.} ``Bonsai: efficient fast failover routing using small arborescences.'' DSN (2019).

\bibitem{Foerster2}Foerster, K.-T. \emph{et al.} ``CASA: congestion and stretch aware static fast rerouting.'' INFOCOM (2019).

\bibitem{LevitinKarpovskyMustafa}Levitin, L., Karpovsky, M., and Mustafa, M. ``Minimal sets of turns for breaking cycles in graphs modeling networks.'' \emph{IEEE Trans. Parallel. Dist. Sys.} {\bf 21}, 1342 (2014).

\bibitem{LiuEtAl}Liu, H. H. \emph{et al.} ``Traffic engineering with forward fault correction.'' SIGCOMM (2014).

\bibitem{PageEtAl}Page, L. \emph{et al.} ``The PageRank citation ranking: bringing order to the web.'' Preprint (1999).

\bibitem{Pignolet}Pignolet, Y.-A., Schmid, S., and Tredan, G. ``Load-optimal local fast rerouting for resilient networks.'' DSN (2017).

\bibitem{Stephens}Stephens, B. and Cox, A. L. ``Deadlock-free local fast failover for arbitrary data center networks.'' INFOCOM (2016).

\bibitem{SucharaEtAl}Suchara, M. \emph{et al.} ``Network architecture for joint failure recovery and traffic engineering.'' SIGMETRICS (2011).

\bibitem{WangEtAl}Wang, Y. \emph{et al.} ``R3: resilient routing reconfiguration.'' SIGCOMM (2010).

\end{thebibliography}
\end{document}